\documentclass[twocolumn]{aastex631}

\usepackage{float}
\usepackage{booktabs}
\usepackage{amsmath}
\usepackage{graphicx}
\usepackage{xcolor}

\begin{document}

\title{Characterization of the Cherenkov Photon Background for Low-Noise Silicon Detectors in Space}

\author[0009-0001-3291-3010]{Manuel E. Gaido}
\affiliation{Department of Physics, University of Chicago, Chicago, IL 60637, USA}
\affiliation{Kavli Institute for Cosmological Physics, University of Chicago, Chicago, IL 60637, USA}
\affiliation{Fermi National Accelerator Laboratory, P.O.\ Box 500, Batavia, IL 60510, USA}
\affiliation{Universidad de Buenos Aires, Buenos Aires, Argentina}
\affiliation{LAMBDA, FCEyN, Buenos Aires, Argentina}

\author[0000-0002-1126-2941]{Javier Tiffenberg}
\affiliation{Fermi National Accelerator Laboratory, P.O.\ Box 500, Batavia, IL 60510, USA}
\affiliation{LAMBDA, FCEyN, Buenos Aires, Argentina}

\author[0000-0001-8251-933X]{Alex Drlica-Wagner}
\affiliation{Fermi National Accelerator Laboratory, P.O.\ Box 500, Batavia, IL 60510, USA}
\affiliation{Kavli Institute for Cosmological Physics, University of Chicago, Chicago, IL 60637, USA}
\affiliation{Department of Astronomy and Astrophysics, University of Chicago, Chicago, IL 60637, USA}
\affiliation{NSF-Simons AI Institute for the Sky (SkAI),172 E. Chestnut St., Chicago, IL 60611, USA}

\author[0000-0003-4238-6813]{Guillermo Fernandez-Moroni}\
\affiliation{Fermi National Accelerator Laboratory, P.O.\ Box 500, Batavia, IL 60510, USA}

\author[0000-0003-2662-6821]{Bernard J. Rauscher}
\affiliation{NASA Goddard Space Flight Center, Greenbelt MD, United States}

\author[0000-0001-5619-1713]{Fernando Chierchie}
\affiliation{Universidad Nacional del Sur, Bahía Blanca, Argentina}

\author[0000-0002-7952-7168]{Darío Rodrigues}
\affiliation{Universidad de Buenos Aires, Buenos Aires, Argentina}
\affiliation{LAMBDA, FCEyN, Buenos Aires, Argentina}

\author{Lucas Giardino}
\affiliation{Universidad de Buenos Aires, Buenos Aires, Argentina}
\affiliation{LAMBDA, FCEyN, Buenos Aires, Argentina}

\author[0000-0002-1527-7956]{Juan Estrada}
\affiliation{Fermi National Accelerator Laboratory, P.O.\ Box 500, Batavia, IL 60510, USA}

\author[0000-0002-8868-3509]{Agustín J.~Lapi}
\affiliation{Department of Astronomy and Astrophysics, University of Chicago, Chicago, IL 60637, USA}
\affiliation{Kavli Institute for Cosmological Physics, University of Chicago, Chicago, IL 60637, USA}
\affiliation{Fermi National Accelerator Laboratory, P.O.\ Box 500, Batavia, IL 60510, USA}

\begin{abstract}

Future space observatories that seek to perform imaging and spectroscopy of faint astronomical sources will require ultra-low-noise detectors that are sensitive over a broad wavelength range. Silicon charge-coupled devices (CCDs), such as EMCCDs, skipper CCDs, multi-amplifier sensing (MAS) CCDs, and single-electron sensitive read out (SiSeRO) CCDs have demonstrated the ability to detect and measure single photons from X-ray energies to near the silicon band gap ($\sim$\,1.1\,$\mu$m), making them candidate technologies for this application. In this context, we study a relatively unexplored source of low-energy background coming from Cherenkov radiation produced by energetic cosmic rays traversing a silicon detector. 
We present a model for Cherenkov photon production and absorption that is calibrated to laboratory data, and we use this model to characterize the residual background rate for ultra-low-noise silicon detectors in space. We study how the Cherenkov background rate depends on detector thickness, variations in solar activity, and the contribution of heavy cosmic ray species ($Z > 2$). We find that for thick silicon detectors, such as those required to achieve high quantum efficiency at long wavelengths, the rate of cosmic-ray-induced Cherenkov photon production is comparable to other detector and astrophysical backgrounds. We apply our Cherenkov background model to simulated spectroscopic observations of extra-solar planets, and we find that thick detectors continue to outperform their thinner counterparts at longer wavelengths despite a larger Cherenkov background rate. Furthermore, we find that minimal masking of cosmic-ray tracks continues to maximize the signal-to-noise ratio of very faint sources despite the existence of extended halos of Cherenkov photons.

\end{abstract}

\section{INTRODUCTION} \label{sec:intro}

Silicon charge-coupled devices (CCDs) have emerged as the most popular detector technology for space-based astronomical observations ranging from soft X-rays to the very near-infrared (NIR; ${\sim}\,1.1\,\mu$m). As sensitivity to faint astronomical sources continues to advance, intrinsic backgrounds in CCD detectors become an increasingly important consideration. For example, the search for potentially habitable Earth-like planets around Sun-like stars is one of the priority areas identified by the most recent Decadal Survey on Astronomy and Astrophysics \citep{Astro2020}. NASA's new flagship mission concept, the {\it Habitable Worlds Observatory}, seeks to perform direct imaging and spectroscopy of Earth-like planets around Sun-like stars \citep{Clery:2023}. Such observations seek to detect ``biosignatures'', aspects of a planet's atmosphere or surface that may indicate the presence of life \citep{Seager:2013}. However, direct spectroscopy of the atmospheres of Earth-like exoplanets presents an enormous technological challenge \citep{Crill:2017,Crill:2022}. The light of the host star greatly outshines the reflected light of the planet, which may have an average rate of several photons per hour \citep{Seager:2010,Rauscher:2019}. In this regime of signal-starved observations and long integration times, ultra-low-noise detectors are vital. In particular, recent technology reports have stated the need for detectors that are sensitive at ultraviolet and visible wavelengths with a readout noise of ${<}\,0.1$\,e$^-$\,rms/pix \citep{Crill:2017,Crill:2022}. Given the precision required for these observations and the limited population of potentially observable exo-Earths \citep{Mamajek:2024}, it is critical to identify and characterize sources of detector, instrumental, environmental, and astrophysical backgrounds in order to accurately estimate the signal-to-noise ratio (SNR) of future exo-planet observations. 

Previous studies have considered several backgrounds that will affect the ability of a future space telescope to directly image and perform spectroscopy of exoplanets \citep{Shaklan:2013, Crill:2017, Rauscher:2019, Crill:2022, Steiger:2024}. These models often include intrinsic detector noise sources (e.g., readout noise, dark current, and clock-induced charge), as well as instrumental and astrophysical backgrounds (e.g., residual speckle light, zodiacal light, and exozodiacal light).
Energetic cosmic rays are another familiar and well-characterized environmental background when operating in the intense radiation environment of space. 
The primary cosmic-ray track can be relatively easily identified through high charge occupancy, which allows for effective masking in science images  \citep[e.g.,][]{vanDokkum:2001}. 
While the cosmic-ray rate sets important limitations on instrument design and operation (e.g., radiation hardness, maximum exposure time, etc.), it is generally assumed to have a negligible impact on the rate of single-photon events that could be confused with faint astronomical signals.
Here, we identify and characterize Cherenkov photon emission from energetic cosmic rays as a previously understudied source of background for photon-starved observations from space.
Cherenkov photons that are emitted with energies near the silicon bandgap can travel a significant distance before depositing their energy in pixels that are outside conventional cosmic-ray masks. 
When combined across multiple exposures, Cherenkov photons represent a background that is indistinguishable from astronomical photons.
In this paper, we demonstrate that for conventional cosmic-ray masking approaches, the Cherenkov photon background is comparable in magnitude to other astrophysical and detector backgrounds that are commonly considered.

This paper is organized as follows. In Section~\ref{sec:cherenkov}, we briefly review the theory of Cherenkov radiation, while in Section~\ref{sec:model} we describe our Cherenkov photon production model.
In Section~\ref{sec:simulation}, we describe our simulations of the cosmic-ray environment and Cherenkov photon production at Earth--Sun Lagrange Point 2 (L2).
In Section~\ref{sec:results}, we discuss the results of our simulations in the context of exoplanet sensitivity estimates, and we conclude in Section~\ref{sec:conclusion}.
This paper improves and extends previous work presented in \citet{Gaido:2024}.

\section{CHERENKOV RADIATION}
\label{sec:cherenkov}

\begin{figure*}[t!]
\begin{center}
\begin{tabular}{c}
\includegraphics[width=0.5\textwidth]{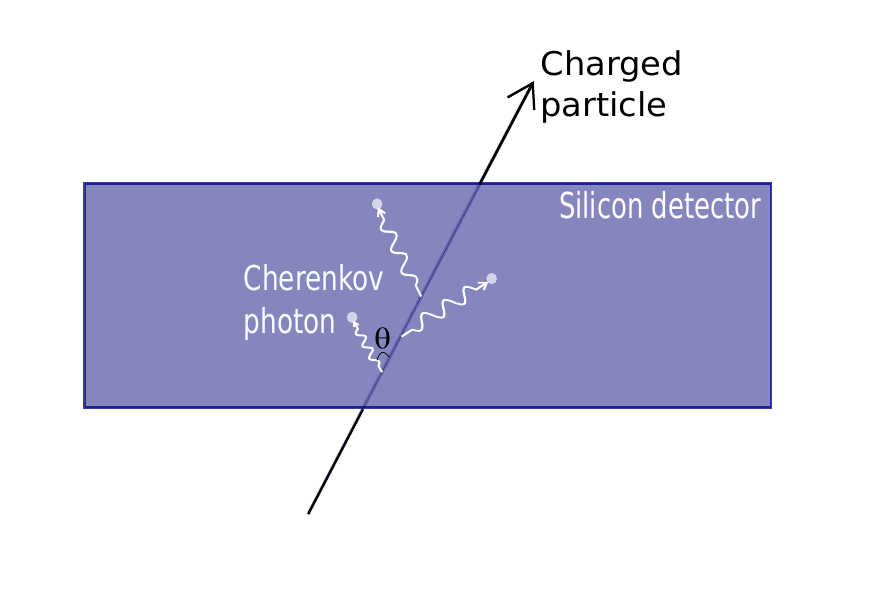}
\hspace{2cm}
\includegraphics[width=0.3\textwidth]{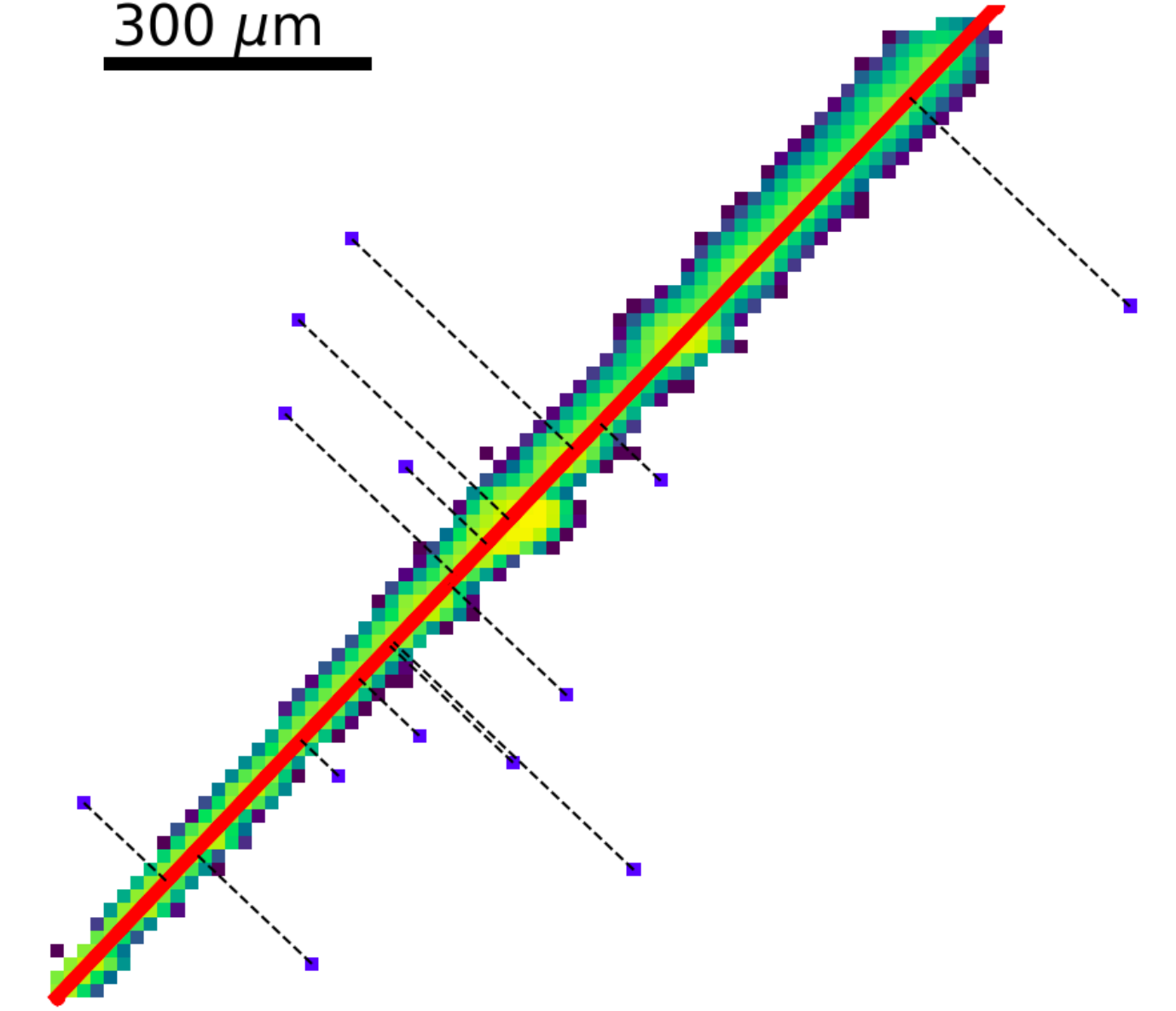}
\end{tabular}
\end{center}
\caption{(Left) Schematic representation of Cherenkov emission in silicon. (Right) Measured muon track with best fit track line (red line) and distances to Cherenkov photons measured orthogonal to the track (black lines).}
\label{fig:schematic-cherenkov}
\end{figure*}

The Cherenkov effect \citep{Cherenkov:1934, Cherenkov:1937} is the process by which a charged particle traveling through a polarizable medium emits electromagnetic radiation (Figure~\ref{fig:schematic-cherenkov}). In order for Cherenkov emission to take place, the charged particle must be traveling faster than the phase velocity of light in the medium. In such cases, the electric dipoles of the molecules in the material are excited, emitting a constructively interfering conical electromagnetic wavefront analogous to a sonic boom \citep{Jelley:1955}. 

\citet{Frank:1937} described the spectrum of electromagnetic radiation that is emitted as a function of particle charge, $q$, and velocity relative to the speed of light in vacuum, $\beta \equiv v/c < 1$, as well as the dielectric function of the medium, $\epsilon \left( \omega \right)$, which depends on the emitted frequency, $\omega$. The differential number of photons produced per unit length per unit frequency is described as \citep{Budini:1953}
\begin{equation}\label{eq:cherenkov}
    \frac{d^2 N_{\gamma}}{dx d \omega} = q^2 \alpha \left(  1 - \frac{\text{Re} \{ \epsilon \left( \omega \right) \} }{\beta^2 | \epsilon \left( \omega \right) |^2} \right), ~~ \beta > \frac{1}{\sqrt{\text{Re} \{ \epsilon \left( \omega \right) \} }},
\end{equation}
where $\alpha$ is the fine-structure constant.
Cherenkov radiation is emitted in a conical wavefront with a vertex at the position of the charged particle and a characteristic Cherenkov angle, $\theta_{c}$, measured between the emitted photon direction and the velocity of the source particle,
\begin{equation}
    \cos (\theta_{c}) = \frac{\sqrt{\text{Re}\{ \epsilon \left( \omega \right)\}}}{\beta \vert \epsilon \left(  \omega \right)\vert}.
\end{equation}\label{eq:theta-ch}

The dielectric function depends on the characteristics of the material, and in silicon there is a sharp cutoff in Cherenkov photon energies at $\sim 4\,{\rm eV}$ \citep{Du:2022}. 
Depending on energy (i.e., the frequency, $\omega$), Cherenkov photons can be promptly absorbed or travel a considerable distance from the charged-particle track.
After Cherenkov photons are emitted, the distance that they travel from the track of the charged particle before being absorbed is set by the photon attenuation length \citep{Du:2022}
\begin{equation}\label{eq:attenuation}
  \ell_\gamma = \frac{1}{\omega \sqrt{ 2 |\epsilon(\omega)| - 2 \text{Re}\{ \epsilon(\omega)}\}}.
\end{equation}
Photons with energies $\gtrsim 2\,{\rm eV}$ have an absorption length in silicon that is smaller than the usual track width ($\lesssim 15\,\mu$m $ = 1\,\text{pix}$), making them invisible in an image taken by a silicon detector such as a CCD. Furthermore, photons with energies $\lesssim$ 1.1\,eV always escape the detector because their energy is less than the silicon bandgap and their absorption length is much larger than the dimensions of the detector.
In the intermediate energy range between $1.2\,{\rm eV} \lesssim E \lesssim 2\,{\rm eV}$, a Cherenkov photon can travel an appreciable distance in silicon before being absorbed to generate a single electron-hole pair via the photoelectric effect, which we refer to as a ``single-electron event'' \citep{Ramanathan_2020}. The production of single-electron events from Cherenkov photons was recently identified as an important background in dark matter direct detection experiments using CCDs \citep{Du:2022}. Cherenkov photon production in CCDs was recently measured in the laboratory \citep{Giardino:2022,moroni2025} and considered as a background for space-based astronomical observations \citep{Gaido:2024}.

\section{CHERENKOV MODEL AND VALIDATION}
\label{sec:model}

To model Cherenkov emission in silicon, we first integrate Eq.~\ref{eq:cherenkov} in frequency space by setting a constant value for the particle velocity, $\beta$. The range of integration is determined by the photon frequencies that satisfy the Cherenkov emission condition for a given velocity (Eq.~\ref{eq:cherenkov}). Due to the nature of the dielectric function of silicon \citep{GREEN20081305, Du:2022}, integration always takes place in a single, continuous frequency interval, 
 
\begin{equation}\label{eq:ngamma}
  \langle N_{\gamma} \rangle = \int_{\text{track}}\int^{\omega_{max}}_{\omega_{min}} \frac{d^2 N_{\gamma}}{dx d \omega} d \omega dx.
\end{equation}

\noindent The value of $\langle N_{\gamma} \rangle$ is the expected number of emitted photons per track trajectory fixing velocity and particle type. In order to simulate Cherenkov photons, we draw from a Poisson distribution with rate $\langle N_{\gamma} \rangle$. The initial position and direction of the emitted photon is determined by drawing a random position uniformly along the particle track, a direction set by the Cherenkov emission angle $\theta_{c}$ (Eq.~\ref{eq:theta-ch}), and a random azimuthal angle drawn from a uniform random distribution, $U(0, 2\pi)$. The energy is drawn from a uniform distribution across the allowed emission frequencies. This approximation is accurate for ultra-relativistic particles that make up the vast majority of cosmic rays responsible of Cherenkov emission, as well as the majority of radiation emitted. Cherenkov photons will propagate a random distance through the silicon detector material before being absorbed (Eq.~\ref{eq:attenuation}). If the photon reaches a detector surface, we assign a 25\% probability of being absorbed and generating an electron-hole pair.

\begin{figure}[t]
\begin{center}
\begin{tabular}{c}
\includegraphics[scale=1.15]{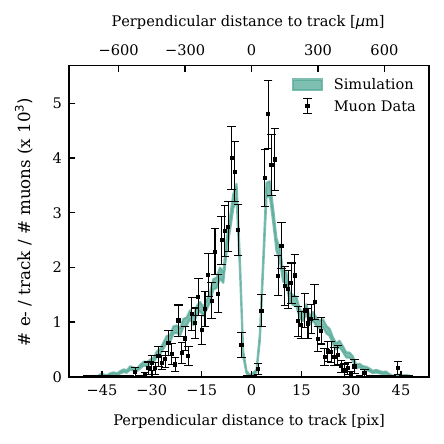}
\end{tabular}
\end{center}
\caption{\label{fig:see-distance-distribution} Distribution of perpendicular distances between muon tracks and single-electron events for images taken by \citet{Giardino:2022} (black points) and simulations from our model (teal band). Simulations include an additional dark current rate estimated from the dataset in order to match the operational conditions. The dip centered at zero comes from the masking applied to energetic particle tracks, which is applied to both the data and simulations.
}
\end{figure}

Our model for producing Cherenkov photons is applied to detailed simulations of cosmic-ray propagation and interaction in silicon detectors performed with the Geant4 simulation toolkit \citep{Geant4:2002}. Geant4 has been used extensively to simulate the interactions of energetic particles with CCD detectors \citep[e.g.,][]{Awaki:2002,Ishiwatari:2006,DAMIC:2016,Oscura:2022,Arnquist:2023}.
Using Geant4, we can simulate cosmic-ray interactions in silicon detectors with high spatial and physical fidelity.  We generate cosmic-ray events by drawing from user-defined particle type, spatial, angular, and energy distributions. The primary particle is propagated through the detector material and each interaction is tracked. When secondary particles are generated they are also tracked. 

We assume no external shielding around the detector in our Geant4 simulations. In reality, shielding by passive material in the instrument and spacecraft would reduce the energy of incoming particles, create secondary showers from interactions of the primary cosmic ray in the shielding material, and change the angular/energy distribution of events. 
Quantitative statements about the effects of shielding on the Cherenkov photon rate would require a detailed model of the instrument and spacecraft, and are thus beyond the scope of the current analysis. 

For each particle, we store each step of its propagation through the silicon detector. A propagation step contains the initial and final position, velocity, deposited energy, and particle type. The charge deposited at each step is diffused in the detector, simulating the drift of charge carriers generated in the bulk of the detector, and its spread is positively correlated with the depth of interaction \citep{PhysRevLett.125.171802}. 
At each propagation step, we use Eq.~\ref{eq:ngamma} to calculate $\langle N_\gamma \rangle$, randomly draw Cherenkov photons as described above, and propagate them by a distance $\ell_\gamma$ (Eq.~\ref{eq:attenuation}). Each photon either escapes the detector or is absorbed to produce a single-electron event.
In this way, we self-consistently simulate images containing charged particle tracks and their corresponding low-energy halos of Cherenkov photons. 

We validate our model against cosmic-ray data collected with an ultra-low-noise silicon skipper CCD capable of counting single-electron events associated with Cherenkov emission. In particular, \citet{Giardino:2022} and \citet{moroni2025} present a laboratory data set of cosmic-ray muons observed with a 675-$\mu$m-thick skipper CCD comprised of 6144$\times$866 15\,$\mu$m pixels. This device was designed at LBNL and fabricated at Teledyne DALSA Semiconductor on high-resistivity silicon. In order to study the low-energy halo around high-energy events, the skipper CCD was operated in ``Smart Skipper'' mode with both energy and region-of-interest readout \citep{Chierchie:2021}. This readout configuration provides single-photon/single-electron resolution around energetic events while minimizing the readout time and dark current. Filtering for atmospheric muons, we can recreate these events in Geant4 by setting their incidence direction as measured in the original images and assuming that their energy is the characteristic atmospheric muon energy \citep[$\sim$ 4\,GeV;][]{Autran:2018}. By simulating several realizations of the same accepted muon events, we can compare the simulated and observed distributions of single-electron events as a function of the perpendicular distance to the muon track. We also measured the dark current event rate from the images and included this in our comparison with simulations. 
We find good agreement between data and simulations (Figure~\ref{fig:see-distance-distribution}), which gives us confidence in extending our model to predict the Cherenkov photon background for silicon detectors in space. 

\begin{figure}[t]
    \centering
    \includegraphics[width=\columnwidth]{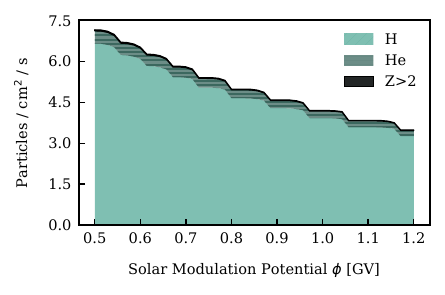}
    \caption{Expected flux of particles as a function of solar activity. Hydrogen nuclei make up the majority of the expected events, followed by helium nuclei. The contribution of heavier nuclei is barely visible, comprising $\sim0.6\%$ of the total event rate. Rates are obtained from the Cosmic-Ray Data Base \citep{Maurin_2023}.}
    \label{fig:composition}
\end{figure}

\section{SIMULATION OF SPACE ENVIRONMENT}
\label{sec:simulation}

We simulate the cosmic-ray environment at L2, the expected operational location for the {\it Nancy Grace Roman Space Telescope} \citep{Spergel:2015, Akeson:2019} and the {\it Habitable Worlds Observatory} \citep{Clery:2023}.
The cosmic-ray environment at L2 closely follows the local interstellar cosmic-ray environment, which is composed primarily of protons with a steeply falling energy spectrum~\citep{hillas2006cosmic, 10.1093/ptep/ptaa104}. 
We obtain the cosmic-ray spectrum at L2 from the Cosmic-Ray Data Base \citep[CRDB;][]{Maurin_2023}, which provides the local interstellar spectrum (LIS) for fully ionized nuclei and includes a model for variation due to solar activity. 
Protons account for more than 90\% of the total cosmic-ray flux at L2, while helium nuclei (alpha particles) are responsible for the majority of the remaining flux \citep{Maurin_2023}. Heavier nuclei ($Z > 2$) are comparatively negligible in flux and contribute little to the Cherenkov photon rate (see Appendix~\ref{app:heavies}). 

Solar activity varies on an approximately 11 yr cycle, modulating the cosmic-ray flux within the heliosphere. This modulation affects both the intensity and the energy distribution of cosmic rays observed at L2. To incorporate solar modulation, we follow the Force-Field approximation, as described in \cite{Zhu_2024}. In this approximation, the effects of solar modulation can be parameterized by the solar modulation potential, $\phi$, which ranges from $0.5 \leq \phi \leq 1.2$\,GV \citep{Maurin_2023}.  This results in a higher particle rate at low solar activity and a lower rate at high solar activity, which is consistent with observations. Our model for the flux and composition of cosmic rays at L2 as a function of solar activity is shown in Figure~\ref{fig:composition}. 
To quantify the effects of varying solar activity, we scale the number of simulated events per exposure according to the modulation shown in Figure~\ref{fig:composition}. Solar activity shifts the minimum expected energy of cosmic rays as well; however, when considering only protons and alpha particles, the change in particle count dominates. As a reference, the minimum proton energy considered across all solar activity levels is 25\,MeV.

\begin{figure}[t]
\begin{center}
\begin{tabular}{c} 
\includegraphics[width=\columnwidth, trim={0 2cm 0 2cm}, clip]{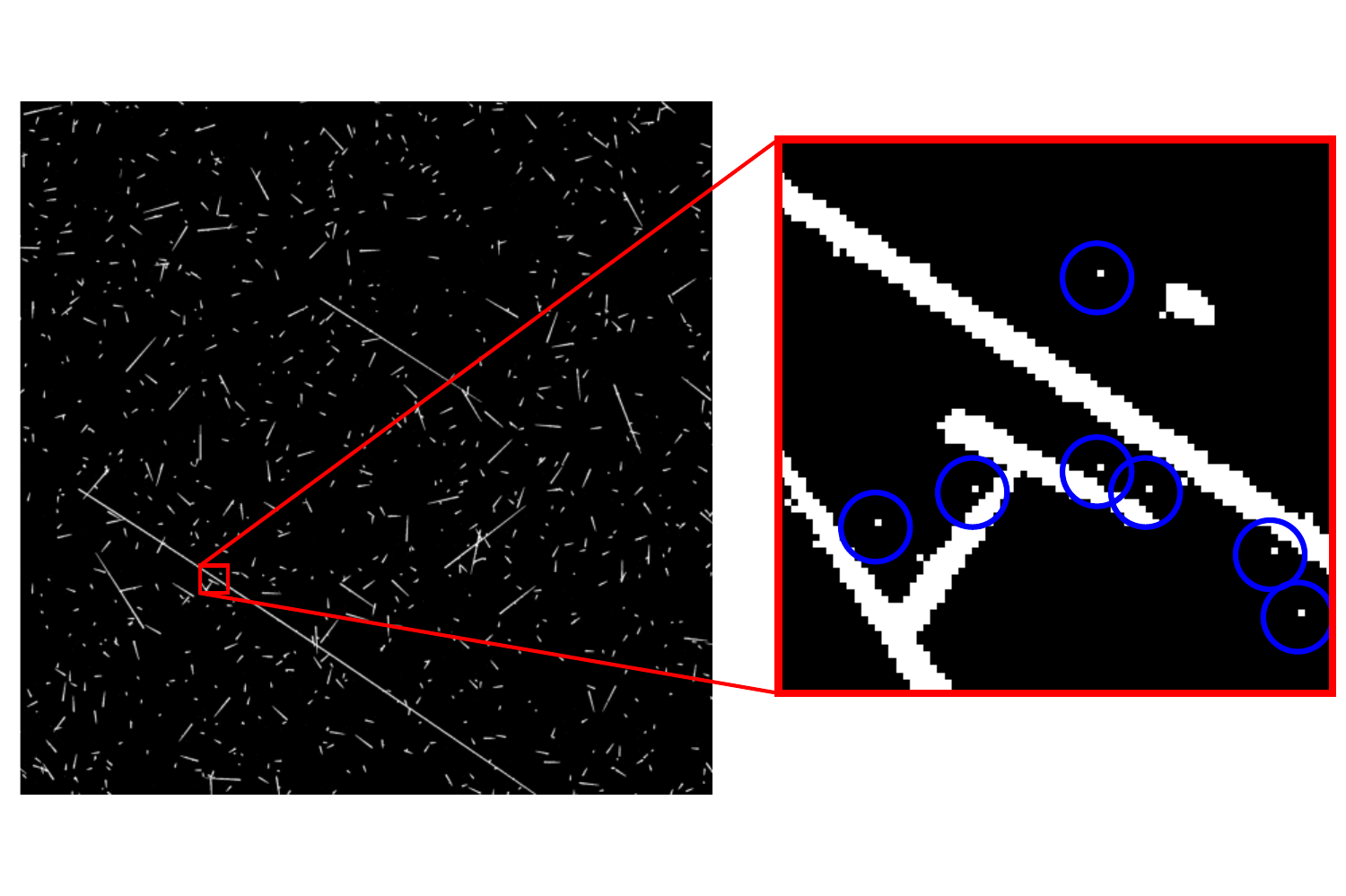}
\end{tabular}
\end{center}
\caption{\label{fig:example-exposure} Illustrative simulation of 30\,s exposure for a 250-$\mu$m-thick silicon detector composed of 2000$\times$2000, $15\,\mu$m pixels. This image includes cosmic-ray proton tracks and secondary Cherenkov photons. The inset shows a small region of the image so that cosmic-ray tracks and Cherenkov photons (circled in blue) can be seen clearly. The cosmic-ray rate roughly corresponds to maximum solar activity.
}
\end{figure}

\begin{figure*}[t]
\begin{center}
\includegraphics[width=0.8\textwidth]{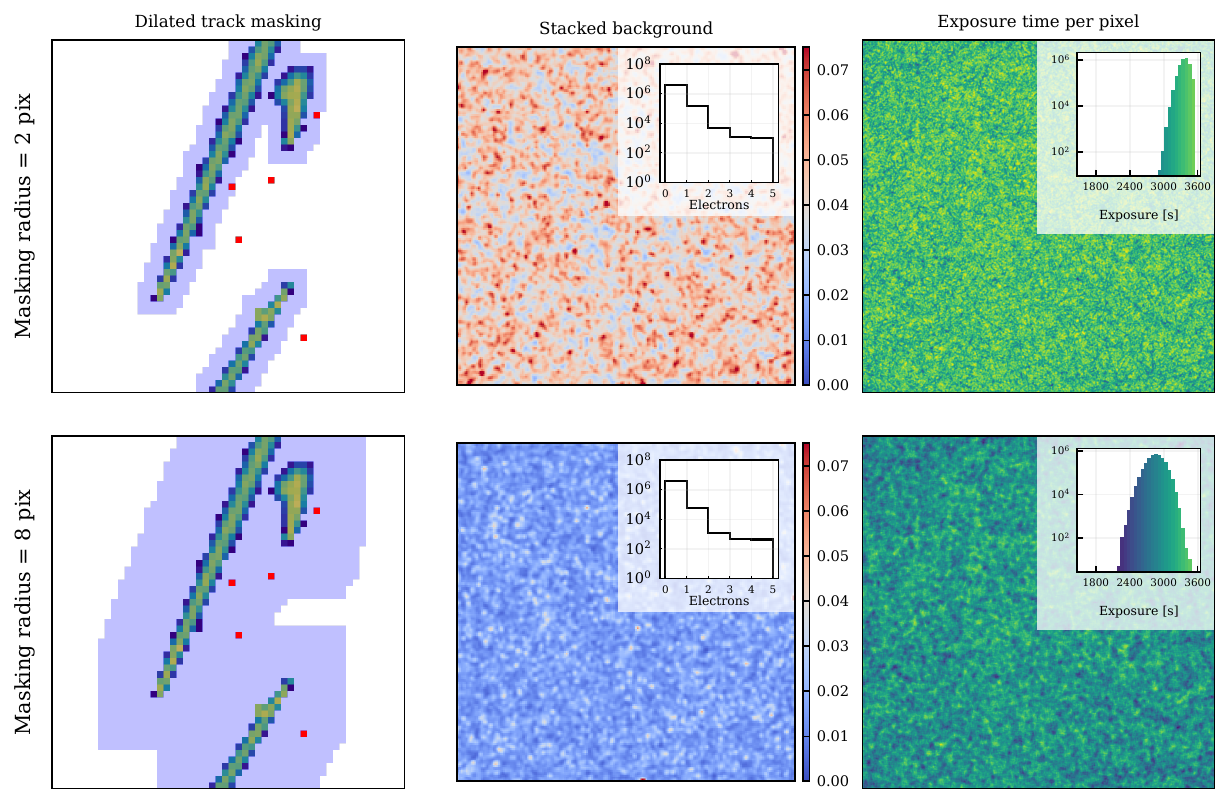}
\end{center}
\caption{(Left) Examples of 2 pixel (top) and 8 pixel (bottom) masking radii (light blue) around simulated proton tracks (green/yellow pixels) and secondary Cherenkov photons (red pixels) for a 250-$\mu$m-thick silicon detector. (Middle) Residual Cherenkov photon rate after masking and combining 120$\times$30\,s exposures, binned in 2$\times$2 blocks and convolved by a Gaussian kernel with a width of 1 pix. Binning and blurring are performed to better visualize the charge distribution. (Right) Effective exposure time for the combination of 120$\times$30\,s exposures after masking.
}
\label{fig:comparative-panel}
\end{figure*}

The silicon detectors are modeled as bulk silicon with a specified thickness and surface area.
The interactions of cosmic rays with the silicon of the detector and the resulting Cherenkov photon production are simulated following the procedure described in Section~\ref{sec:model}.
Primary cosmic rays are simulated uniformly on the surface of a sphere surrounding the detector yielding an isotropic angular distribution and a uniform spatial distribution.
No surrounding material (e.g., instrument or spacecraft) is modeled. Minimal shielding (e.g., $\gtrsim$\,10\,mm of aluminum equivalent) will dominantly affect low-energy events and is only expected to change the high-energy proton flux by $\sim$\,20\% at energies $\gtrsim$\,25\,MeV \citep[e.g.,][]{Dawson:2008}. Considering this, the rates we present here are slightly higher than those measured for the {\it James Webb Space Telescope} operating at L2 (B.~J.~Rauscher \& D.~J.~Fixsen, in prep.). 
We use Geant4 to generate a sufficient number of cosmic-ray events, such that each event is used only once per simulated set of exposures.

To emulate a realistic observing strategy, we divide the simulated data into exposures of 30\,s each and combine 120 exposures to simulate 1 hour of observation time.
For each exposure, we include the cosmic rays that interact with the detector during readout, where the readout time is specified as a configurable detector parameter.
We augment our sample of simulated exposures by performing rotations and reflections to produce eight unique image orientations. We randomly draw sets of short exposures using bootstrap resampling to estimate the statistical scatter of our results for long observations. We find that the statistical scatter is negligible compared to the effect of solar modulation.
Figure \ref{fig:example-exposure} shows an example of a simulated 30\,s image for a 250-$\mu$m-thick silicon detector with instantaneous readout.

Cosmic-ray tracks are identified in each exposure as pixels that contain $\geq 50$\,e$^-$ and these pixels are masked. This mask is then dilated by a fixed number of pixels (i.e., the ``masking radius''). Exposures are stacked after masking high-energy tracks to obtain the combined observation.
Pixels in the combined observation have different effective exposure times due to masking, and Cherenkov photons that reside outside the track masks contribute a residual background rate.
For a single image, the residual unmasked Cherenkov photons and masked exposure time clearly trace the cosmic-ray tracks; however, no coherent structure is visible in the combined observation (Figure~\ref{fig:comparative-panel}). 
The left panel of Figure~\ref{fig:comparative-panel} shows the difference between two masking radii. The middle panel shows the spatial distribution of residual Cherenkov events in the combined image after masking regions around high-energy events. The smaller masking radius (top) produces a higher mean residual background rate and correspondingly higher shot-noise fluctuations, while the larger masking radius (bottom) results in a lower Cherenkov background rate with smaller variations. 
The variation in the effective exposure time per pixel (i.e., the sum of the exposure time when the pixel is unmasked) is shown in the right panel. 
To compute the residual Cherenkov background rate, we count the number of electrons remaining in the combined, processed image and divide by the nominal exposure time of 1 hour.

\section{RESULTS}
\label{sec:results}

\begin{deluxetable*}{lccccc}
\tablecaption{Detector parameters for EMCCD and MAS CCD models.\label{tab:detectors}}
\tablehead{
\colhead{Parameter} &
\colhead{EMCCD} &
\colhead{MAS CCD} & 
\colhead{Units} &
\colhead{References}
}
\startdata
Format &  $1024\times1024$ & $2000\times2000$ & pix & \ldots \\
Pixel Size & $13$ & $15$ & $\mu$m & \ldots  \\
Thickness  & $13$ & $250$ & $\mu$m & \dots  \\
Dark Current  & $4.6\times10^{-4}$ & $6.8 \times 10^{-9}$ & e$^{-}$/pix/sec & (1) (2) \\
Clock-Induced Charge & $0.01$ & $1.5\times10^{-4}$ & e$^{-}$/pix/frame & (1) (2) \\
Readout Noise & $1.7 \times 10^{-6}$ & $2.7\times10^{-4}$ & e$^{-}$/pix/frame & (1) (3) \\ 
Readout Time & 0 & 30 & sec & \ldots  \\
Quantum Efficiency (940\,nm) & 12\% & 89\% & \ldots & (1) (4)  \\
\hline
Cherenkov Rate (2 pix) & $9.2 \times 10^{-7}$ & $1.6\times10^{-5}$ & e$^{-}$/pix/sec & This work  \\
Cherenkov Rate (10 pix) & $2.9 \times 10^{-7}$ & $3.2 \times 10^{-6}$ & e$^{-}$/pix/sec & This work  \\
\enddata
\tablecomments{Cherenkov rates assume cosmic-ray flux at L2 during median solar activity and the specified masking radius. The Cherenkov rate for the MAS CCD includes Cherenkov photons generated during the 30\,s readout time. For the EMCCD, we use the parameter values presented in \cite{Lacy:2019}.}
\tablerefs{(1) \citet{Lacy:2019} (2) \citet{Barak:2022},  (3) \citet{Cancelo:2021}, (4) \cite{Marrufo:2024}.}
\end{deluxetable*}

We simulate the cosmic-ray environment at L2 for two detector configurations (Table~\ref{tab:detectors}). The first configuration mimics the Teledyne-e2v CCD201-20 EMCCD detectors that were initially baselined for the {\it Roman} Coronagraph Instrument \citep[CGI;][]{Spergel:2015, Harding:2016}. We note that for these simulations, we care only about detector thickness and pixel size, and thus our results should be applicable to newer EMCCD detector architectures \citep[i.e., the CCD301, CCD302, and CCD311;][]{Morrissey:2023}. These are relatively thin detectors (13\,$\mu$m thick) with fast readout and good quantum efficiency in the visible. The detector characteristics including readout noise, clock-induced charge, and dark current were collected by \citet{Lacy:2019} based on measurements from \citet{Harding:2016}. We chose these detectors because they were used for the exoplanet sensitivity study presented in \citet{Lacy:2019}.

The second detector configuration corresponds to a future red-optimized silicon detector with a thickness of $250\,\mu$m. This detector configuration is envisioned as a future iteration of the $p$-channel MAS CCD \citep{Holland:2023, Botti:2024}, which inherits the radiation tolerance, low readout noise, and quantum efficiency of the skipper CCD \citep[e.g.,][]{Dawson:2008, Tiffenberg:2017, Marrufo:2024, Roach:2024, Cervantes-Vergara:2025}. The MAS CCD achieves deeply sub-electron readout noise through the repeated measurement of the charge in each pixel by a series of floating gate skipper amplifiers.
We envision a detector with $2000 \times 2000$, $15\,\mu$m pixels read through two serial registers, each containing 300 non-destructive amplifiers with an inter-amplifier separation of 3 pixels.
Such a device could be read in $<$\,30\,s in frame-transfer mode with a readout speed of $\sim70$\,kpix/s \citep{Lapi:2025,Lin:2025}.
Assuming a single-sample readout noise of 2.5\,e$^-$\,rms/pix \citep{Cancelo:2021}, this would result in a final readout noise of $\sim0.14$\,e$^-$\,rms/pix.
We converted the readout noise into the rate of readout-noise-induced spurious charge reported in Table~\ref{tab:detectors}. We note that this spurious charge rate comes from sampling the exponential tail of the readout noise distribution and is very sensitive to the exact value of the readout noise. 
This rate could be reduced by improving the single-sample noise, adding more output amplifiers, or measuring the charge multiple times per amplifier \citep{Lapi:2025}.
Similar readout noise could also be achieved over smaller regions-of-interest in a comparable readout time using fewer amplifiers \citep[][]{Chierchie:2021}.
Furthermore, a thick, $p$-channel CCD read with multiple SiSeRO amplifiers could achieve similar readout noise and speed with fewer output channels \citep{Sofo-Haro:2024}.
We take values for clock-induced charge and dark current from \citet{Barak:2022}; however, this performance has yet to be demonstrated in an astronomy-optimized detector.

\subsection{Detector Thickness}
\label{sec:thickness}

\begin{figure*}[t]
    \centering
    \includegraphics[scale=1.10]{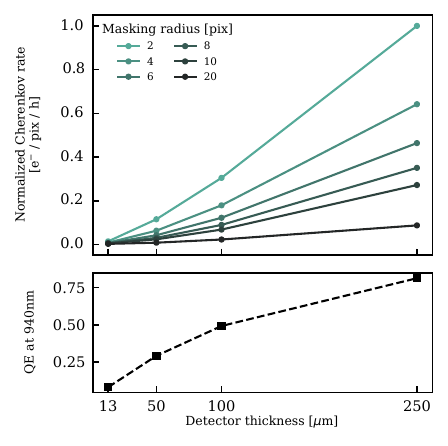}
    \includegraphics[scale=0.55]{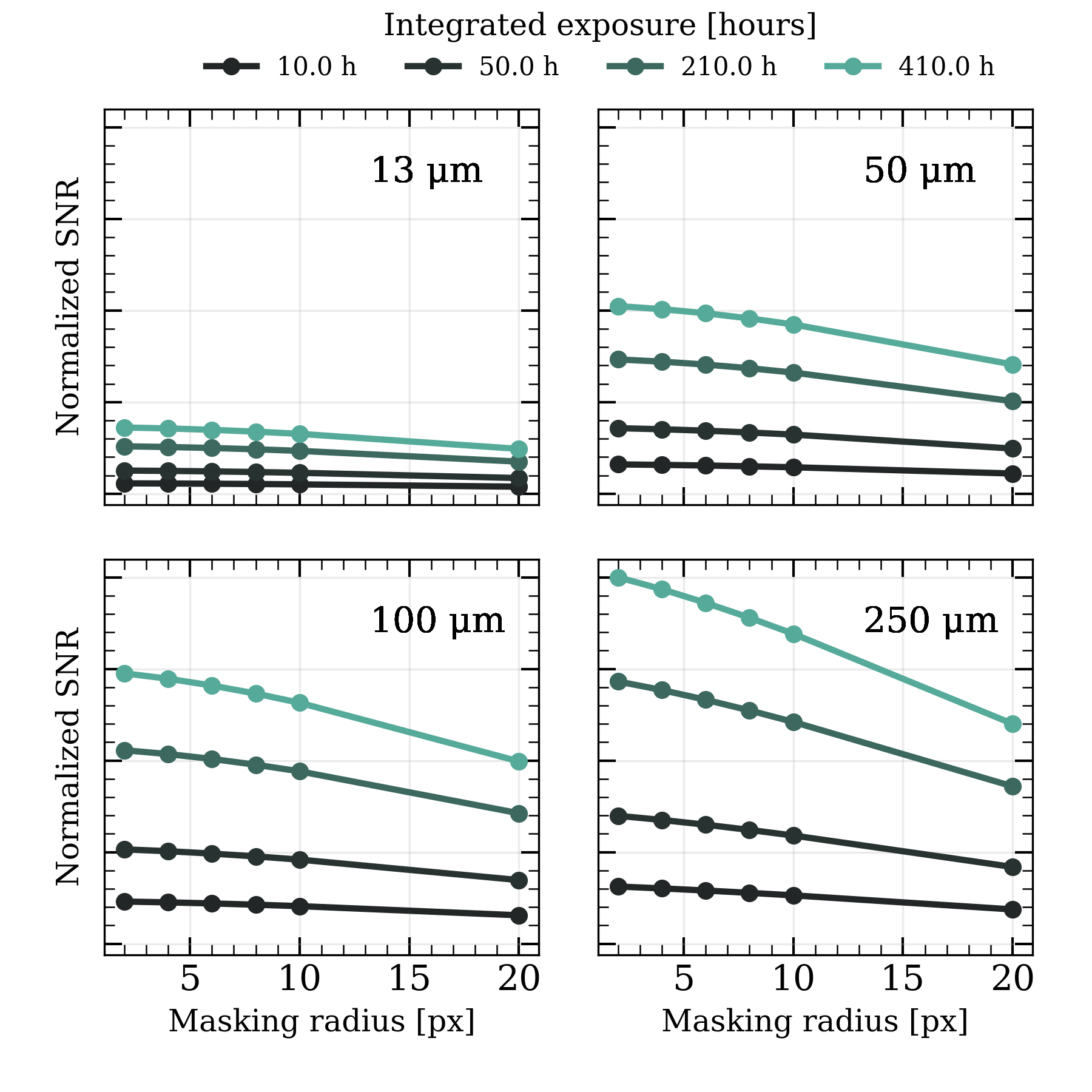}
    \caption{
    (Top left): Cherenkov background rate as a function of detector thickness for different masking radii. (Bottom left): Quantum efficiency (as given by our model of absorption of photons in silicon) at the 940\,nm feature, as a function of detector thickness. (Right): Signal-to-noise ratios at 940\,nm for different detector thicknesses including the residual Cherenkov background rate after masking along with intrinsic detector backgrounds and signal rate from \citet{Lacy:2019}. Thicker detectors are found to have larger signal-to-noise despite having the larger residual Cherenkov background rates due to their enhanced quantum efficiency.
    }
    \label{fig:thickness-scan}
\end{figure*}

The residual Cherenkov background rate depends on detector thickness. Thinner detectors reduce Cherenkov emission by limiting the amount of material traversed by high-energy particles leading to shorter tracks, fewer Cherenkov photons, and a lower probability that Cherenkov photons are absorbed by the detector. However, this suppression comes at the cost of reduced quantum efficiency at longer wavelengths. This trade-off is critical when targeting spectral features such as the $\mathrm{H_2O}$ absorption band at 940\,nm, which is of particular interest in the search for biosignatures \citep[e.g.,][]{Seager:2013,Rauscher:2019}. 

In order to isolate the effect of detector thickness, we start from the EMCCD configuration in Table~\ref{tab:detectors} and artificially change its thickness while keeping all other detector parameters fixed.
For each detector thickness, we adjust the corresponding quantum efficiency and effective exposure time after masking.
Figure~\ref{fig:thickness-scan} shows how the residual Cherenkov background and quantum efficiency change with detector thickness and masking radius.
We compute a simple SNR considering literature values for intrinsic detector backgrounds (Table~\ref{tab:detectors}) and a nominal signal rate of 0.0016\,e$^-$/pix/s from \citet{Lacy:2019}.
We find higher SNR for thicker detectors, indicating that the increased quantum efficiency dominates over the increased residual Cherenkov background. 
Furthermore, our analysis suggests that a smaller masking radius maximizes SNR because a larger masking radius reduces the total effective exposure time. 
We note that the upper bound of a 20\,pix masking radius is much larger than conventional cosmic-ray masking algorithms \citep[e.g.,][]{vanDokkum:2001} and leads to a significant ($> 70\%$) reduction in total signal integration time.

\subsection{Exoplanet Sensitivity}
\label{sec:sensitivity}

We proceed to include our model for the residual Cherenkov photon background in the {\it Roman} CGI exoplanet atmosphere sensitivity model described by \citet{Lacy:2019}.
\citet{Lacy:2019} provide a detailed description and open source code\footnote{\url{https://github.com/blacy/direct-imaging-sims}} for estimating signal and background sources for exoplanet spectroscopy, with a focus on the giant exoplanet observations with the then-current design of the {\it Roman} CGI.\footnote{Since the analysis of Lacy et al. (2019), the Roman-CGI design moved to a slit-prism spectrograph. This reduces the dark current contribution due to a decrease in the number of pixels covered by the signal.}
The \citet{Lacy:2019} model includes intrinsic detector backgrounds (dark current, clock-induced charge, and readout noise), as well as speckle, zodiacal, and exozodiacal light.

\begin{figure*}[t]
\centering 
\includegraphics[width=0.48\textwidth]{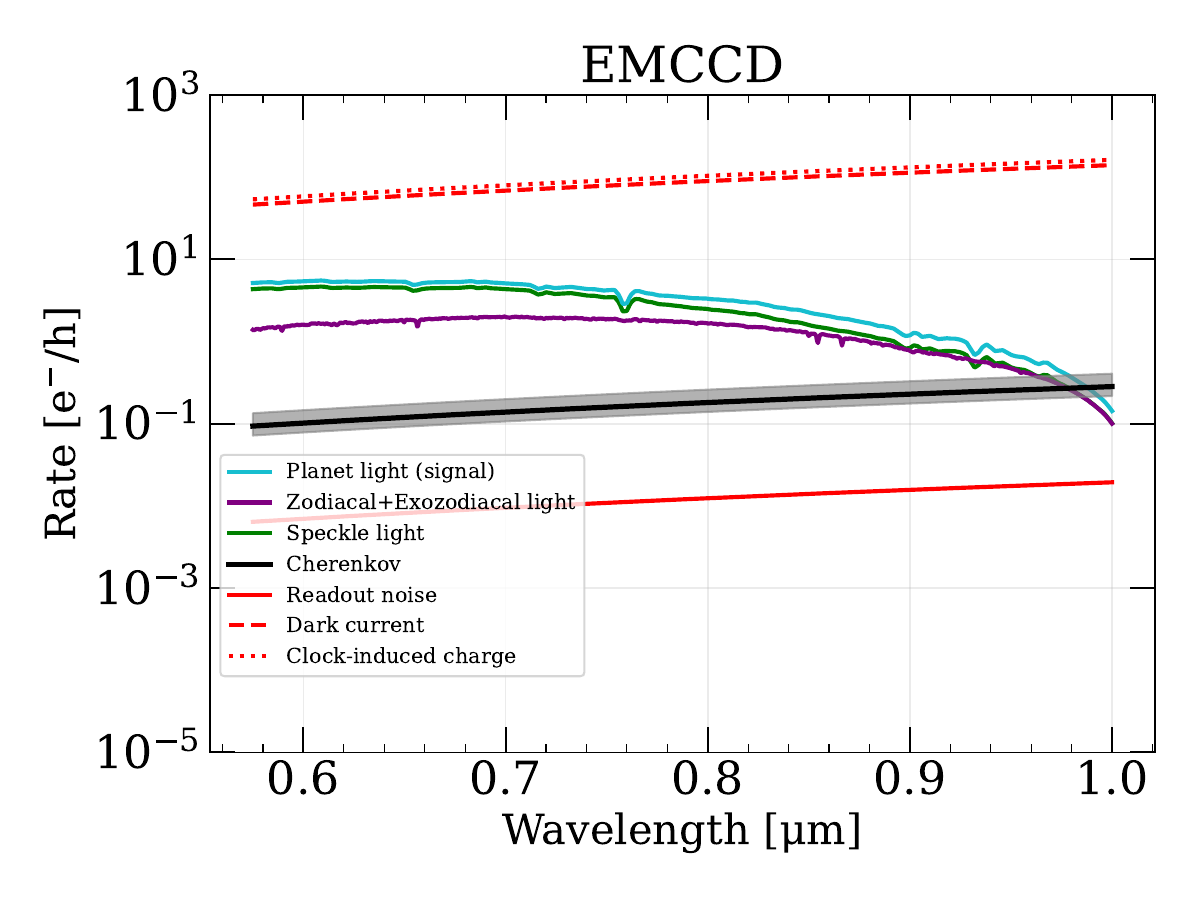}
\includegraphics[width=0.48\textwidth]{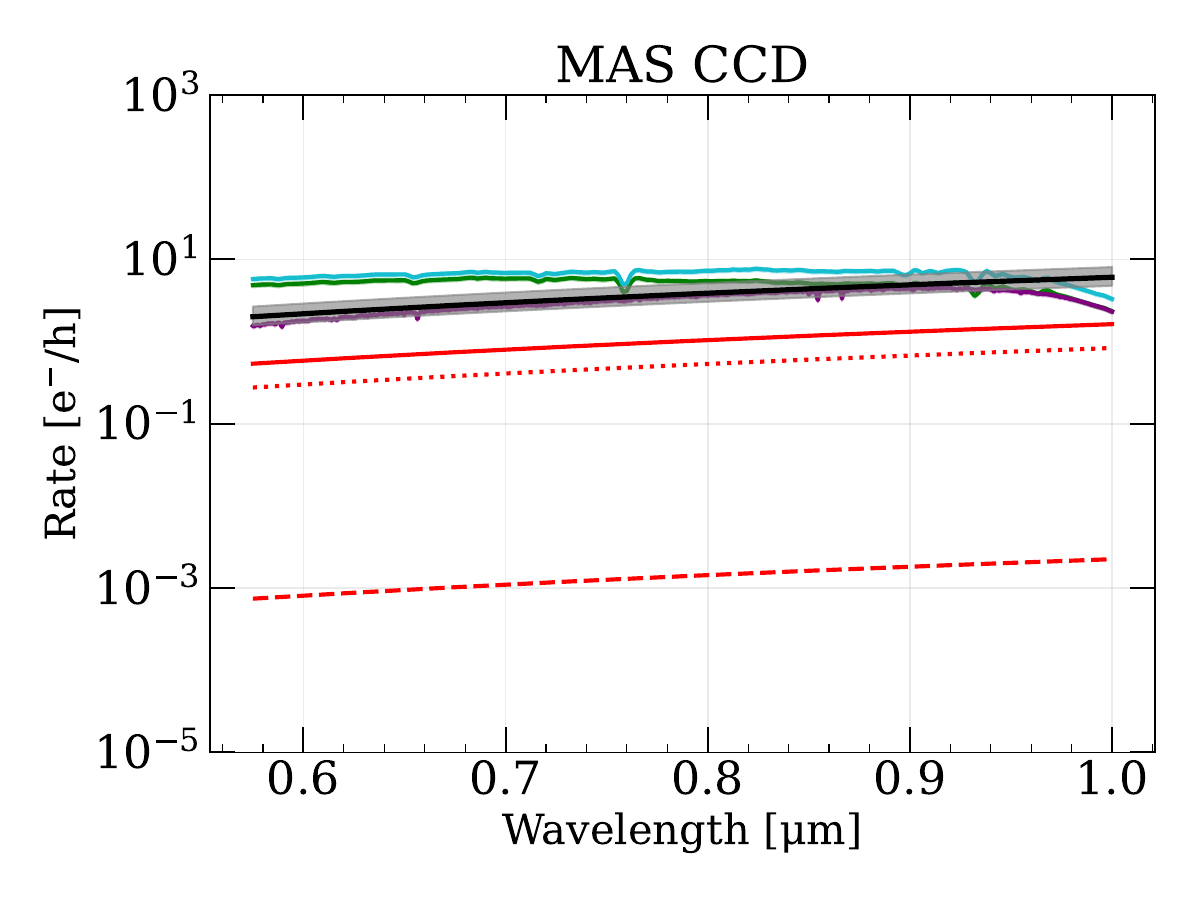}
\caption{Signal and background rates for direct spectroscopy of a Jupiter-radius planet orbiting a 5\,mag G0V star at a separation of 3.8\,AU observed at a distance of 10\,pc using a model of the {\it Roman}-CGI in IFS mode from \citet{Lacy:2019}. Rates are calculated per spectral pixel and the wavelength dependence in the intrinsic detector background rates comes from the changing number of detector pixels per spectral pixel (see \citealt{Lacy:2019}). (Left) The residual Cherenkov background rate is subdominant to the astronomical signal and other backgrounds for a thin EMCCD. (Right) The residual Cherenkov background rate is comparable to the astronomical signal and other backgrounds for a thick, red-sensitive MAS CCD.
}
\label{fig:roman-model}
\end{figure*}

In Figure~\ref{fig:roman-model}, we present estimates for the signal and background rates for the {\it Roman} CGI operating in integral field spectrograph (IFS) mode for an observation of a Jupiter-radius planet orbiting a 5.0\,mag G0V star at a separation of 3.8\,AU observed at a distance of 10\,pc.
Using the \citet{Lacy:2019} model for a 13-$\mu$m-thick Teledyne-e2v CCD201-20 EMCCD, we find that the residual Cherenkov background rate is subdominant to other detector and astronomical backgrounds, as well as the exoplanet signal.
However, this thin detector provides minimal sensitivity at longer wavelengths, with ${\rm SNR} < 1$ at 940\,nm after 410 hours of integrated exposure time.
In contrast, the envisioned 250-$\mu$m-thick fully depleted MAS CCD has a Cherenkov background rate that is larger than other intrinsic detector backgrounds and is comparable to the exoplanet signal and astronomical backgrounds. This configuration yields ${\rm SNR} \gtrsim 10$ at 940\,nm after 410 hours of integrated exposure time. We note that assuming a 30\,s readout time per exposure in the MAS CCD configuration doubles the dark current and residual Cherenkov contributions relative to the signal.
A 2-pixel masking radius is assumed for the residual Cherenkov background rate presented in Figure~\ref{fig:roman-model}, while the gray band represents the variation expected from solar modulation. 
Variations in solar activity affect both the residual Cherenkov background rate and the fraction of the detector that is masked due to cosmic-ray tracks.

\section{CONCLUSIONS}
\label{sec:conclusion}

Combining our model of Cherenkov photon production and absorption (Section~\ref{sec:cherenkov}) with a model for the cosmic-ray environment at L2 (Section~\ref{sec:simulation}), we calculate the single-electron background rate generated by cosmic-ray-induced Cherenkov photons. These photons can travel an appreciable distance from the primary cosmic-ray track and represent an additional instrumental background that has been omitted from previous studies of single-photon sensitive silicon detectors in space. The residual Cherenkov background will prove important for assessing the sensitivity of future instruments that seek to detect very faint astrophysical signals in the intense space radiation environment.

We find that the residual background from cosmic-ray-induced Cherenkov photons can exceed 1\,e$^-$/pix/h in a 250-$\mu$m-thick detector (Table~\ref{tab:detectors}). 
This residual Cherenkov background can be mitigated through aggressive masking around cosmic-ray tracks; however, such an approach results in a loss of effective exposure time per pixel.
Other conventional approaches to transient background removal, such as median stacking or difference imaging, will not work when searching for very faint signals among a background of single-electron events because both the signal and background are stochastic with expectation values of $\ll 1$ event per image.

By analyzing several detector thicknesses, we conclude that thicker detectors provide a higher SNR at long wavelengths (e.g., $\sim 940$\,nm), despite being subject to a larger residual Cherenkov background rate. 
Reduction of the effective exposure time per pixel due to masking is the dominant effect when determining variations in SNR for such observations.
For thicker detectors, we evaluate the impact of heavy cosmic-ray species ($Z > 2$). We determine that, due to their comparatively low flux, heavy species do not contribute significantly to the residual Cherenkov background (Appendix~\ref{app:heavies}). 
Solar modulation changes the flux of cosmic rays and the residual Cherenkov rate for astronomical detectors operating in space. 
This changes the sensitivity to faint astronomical sources by modulating the area of the detector masked by cosmic-ray tracks and the residual Cherenkov background rate.

We emphasize that Cherenkov radiation is an irreducible source of background for \emph{all} silicon detectors in space, as well as any other detectors using dielectric materials where Cherenkov radiation can be emitted and detected \citep{Du:2022}.  
For silicon detectors with coarse timing resolution (e.g., EMCCDs, skipper CCDs, MAS CCDs, and SiSeRO CCDs), Cherenkov photons in the energy range of $1.2 \lesssim E \lesssim 2$\,eV are indistinguishable from photons from astronomical sources, and can only be identified probabilistically due to their higher occurrence rate close to cosmic-ray particle tracks.
The cosmic-ray track length, and thus the Cherenkov photon rate, can be reduced with thinner detectors; however, thin detectors are less sensitive at long wavelengths, which is a critical regime for identifying biosignatures in the spectra of exoplanet atmospheres \citep{Rauscher:2022}.
The residual cosmic-ray-induced Cherenkov photon background is thus an important component to include when investigating the trade space between detector thickness, readout noise, and readout speed for future space missions.\\[-0.5em]

\noindent {\it Software}: Geant4 \citep{Geant4:2002}, matplotlib \citep{Hunter:2007}, numpy \citep{Harris:2020}, scipy \citep{Virtanen:2020}.

\section*{Acknowledgments}
We thank David Maurin for helpful discussions about the cosmic-ray environment at L2. We thank Brianna Lacy for a useful discussion about the {\it Roman} CGI instrument model.
This work was partially funded by NASA APRA (No.\ 80NSSC22K1411), Fermilab LDRD (2019.011, 2022.054), and the Heising-Simons Foundation (\#2023-4611).

The fully depleted skipper CCD was developed at Lawrence Berkeley National Laboratory. CCD development work was supported in part by the Director, Office of Science, of the U.S.\ Department of Energy under No.~DE-AC02-05CH11231. The multi-amplifier sensing (MAS) CCD was developed as a collaborative endeavor between Lawrence Berkeley National Laboratory and Fermi National Accelerator Laboratory. Funding for the design and fabrication of the MAS device described in this work came from a combination of sources including the DOE Quantum Information Science (QIS) initiative, the DOE Early Career Research Program, and the Laboratory Directed Research and Development Program at Fermi National Accelerator Laboratory under Contract No.~DE-AC02-07CH11359.

This document was prepared using the resources of the Fermi National Accelerator Laboratory (Fermilab), a U.S. Department of Energy, Office of Science, Office of High Energy Physics HEP User Facility. Fermilab is managed by Fermi Forward Discovery Group, LLC, acting under Contract No.\ 89243024CSC000002.
The United States Government retains and the publisher, by accepting the article for publication, acknowledges that the United States Government retains a non-exclusive, paid-up, irrevocable, world-wide license to publish or reproduce the published form of this manuscript, or allow others to do so, for United States Government purposes.

\appendix
\section{Contribution of Heavy Nuclei}
\label{app:heavies}

To quantify the contribution of different cosmic-ray species  to the Cherenkov background, we generate separate image sets for hydrogen (H), helium (He), and heavier nuclei ($Z > 2$). Each set of images is generated and processed independently using the same simulation and analysis pipeline. We then compute the background rate expected for a total integration time of one hour, assuming a sequence of $120 \times 30$\,s exposures with instantaneous readout. We evaluate the contribution of each particle species individually, as well as their combined effect.

\begin{figure}[h]
    \centering
    \includegraphics[width=0.6\textwidth]{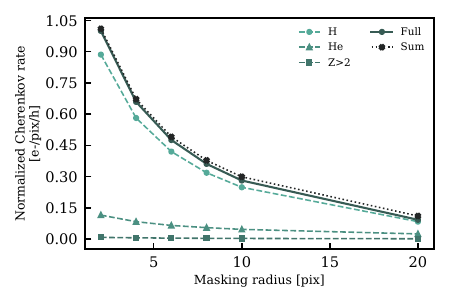}
    \caption{\label{fig:composite-rate} Relative Cherenkov background rate for maximum solar activity (minimum number of particles per exposure) considering cosmic-ray species individually and in combination. Because of track overlap, it is expected that the raw addition of the rates yields a higher value than the rate obtained from the combined images. Curves are normalized to the maximum of the combined rate (``Full'' curve).}
\end{figure}

As noted before, heavy nuclei make a negligible contribution to the overall cosmic-ray rate. While it is expected that heavy nuclei will also contribute negligibly to the Cherenkov photon rate, the correspondingly larger electric charge of heavy nuclei means that more Cherenkov photons are produced per cosmic-ray event (i.e., the $q^2$ dependence in Eq.~\ref{eq:cherenkov}).
From our full simulations, we see that heavy nuclei indeed make a minimal contribution to the residual Cherenkov background rate (Figure~\ref{fig:composite-rate}). This does not mean that heavy nuclei are unimportant to observations; their tracks occupy more pixels than light nuclei, reducing the effective exposure time per pixel after processing.
However, this is suppressed for longer exposure times, since protons dominate the detector occupation and overlap between tracks becomes more likely to occur. 
Thus, we conclude that heavy nuclei can be ignored when estimating the Cherenkov background rate and detector masking; however, the greater charge and more complex event morphologies of heavy nuclei can affect image calibration pipelines \citep[e.g.,][]{Regan:2024}.

\section{Cherenkov Rates}

\begin{figure}[h]
    \centering
    \includegraphics[width=0.75\textwidth]{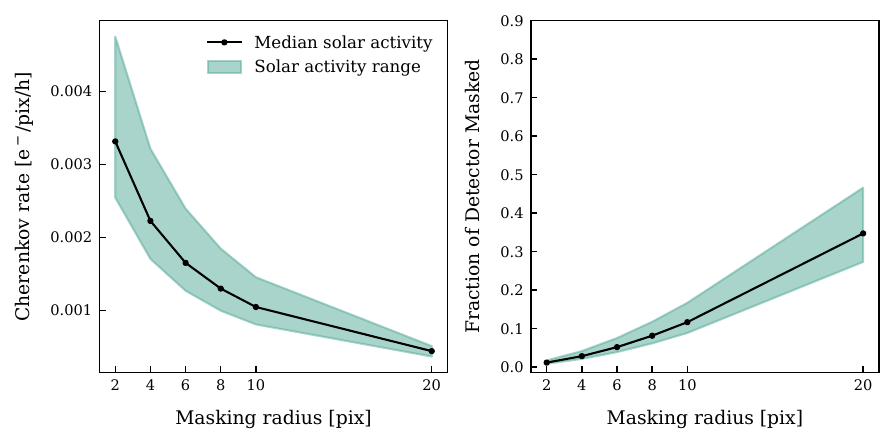}
    \caption{\label{fig:emccd-rate-band} (Left) Residual Cherenkov rate at L2 for the thin EMCCD configuration as a function of masking radius. The shaded band represents the spread due to variations in particle flux corresponding to the variation in the solar modulation potential shown in Figure~\ref{fig:composition}. (Right) Fraction of detector pixels masked due to cosmic ray hits as a function of masking radius.}
\end{figure}

\begin{figure}[h]
    \centering
    \includegraphics[width=0.75\textwidth]{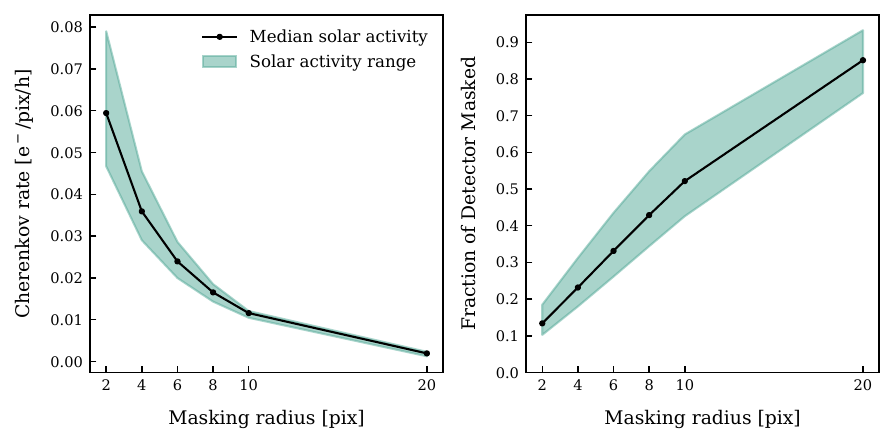}
    \caption{\label{fig:skipper-rate-band} (Left) Residual Cherenkov rate at L2 for the MAS CCD configuration as a function of masking radius. The shaded band represents the spread due to variations in particle flux from changing solar activity, corresponding to the variation shown in Figure~\ref{fig:composition}. (Right) Fraction of detector pixels masked due to cosmic ray hits as a function of masking radius.}
\end{figure}

Considering protons and alpha particles, we present the Cherenkov rates as a function of track masking radius, along with their spread due to varying levels of solar activity. The Cherenkov rates for the EMCCD and MAS CCD configurations are shown in Figures~\ref{fig:emccd-rate-band} and \ref{fig:skipper-rate-band}.
We emphasize that, like the dark current background, this background accumulates during both exposure and readout. For example, in the case of the MAS CCD, we consider a non-zero readout time, which contributes to the observed background even outside of the exposure window. This also affects the fraction of the detector area masked in each exposure due to cosmic-ray hits, thereby reducing the effective exposure time. Additionally, the Cherenkov rate (and masking fraction) depends on how the total observation time is partitioned. In this study, we assume 1 hour of observation divided into 120 exposures of 30\,s each. As a limiting case, if we instead considered a single continuous one-hour exposure, cosmic rays would effectively cover the entire detector.

We note that the rates presented here differ from those calculated in \citet{Gaido:2024} due to changes in the cosmic-ray proton rate, how readout time was factored into the observation time, and how the per-frame rate was computed. 
We believe that the rates presented here represent a more accurate estimate of the residual Cherenkov background rate expected in L2.

\vspace{10em}
\pagebreak
\bibliography{PASPsample631}{}
\bibliographystyle{aasjournal}

\end{document}